\documentclass[12pt,preprint]{aastex}

\begin{document}

\title{GRB 090618: different pulse temporal and spectral characteristics
within a burst}
\shorttitle{GRB 090618: different pulse characteristics within a burst}
\shortauthors{Zhang}

\author{Fu-Wen Zhang}
\affil{College of Science, Guilin University of
Technology, Guilin, Guangxi 541004, China}

\email{fwzhang@pmo.ac.cn}

\begin{abstract}
GRB 090618 was simultaneously detected by
Swift-BAT and Fermi-GBM. Its light curve shows two emission episodes
consisting of four prominent pulses. The pulse in the first episode (episode A) has a smoother
morphology than the three pulses in the second episode (episode B). Using the
pulse peak-fit method, we have performed a detailed analysis of the temporal and spectral characteristics of these four pulses and found out that the first pulse (pulse A) exhibits distinctly different properties than the others in episode B (pulses B1, B2
and B3) in the following aspects. (i) Both the pulse width ($w$) and the
rise-to-decay ratio of pulse ($r/d$, pulse asymmetry) in GRB 090618 are found to be
energy-dependent. The indices of the power-law correlation between $w$ and $E$ for the pulses in episode
B however are larger than that in episode A. Moreover the pulses B1, B2 and B3 tend to be more symmetric
at the higher energy bands while the pulse A displays a
reverse trend. (ii) Pulse A shows a hard-to-soft spectral evolution
pattern, while the three pulses in the episode B follow the light curve trend.
(iii) Pulse A has a longer lag than the pulses B1, B2 and B3. The mechanism which causes the
different pulse characteristics within one single GRB is unclear.
\end{abstract}



\section{Introduction}
Gamma-ray bursts (GRB) have remained enigmatic since their discovery in the late 1960s (for reviews, see Piran 2004; Zhang 2007). Although in the last ten years
our understanding of GRBs has been advanced significantly, due mainly to the
study of GRB afterglows (e.g., Sari et al. 1998; Fan \& Wei 2005; Zhang et al. 2006), the exact mechanism which produces the
prompt gamma-ray emission has not been definitively established
(e.g., Fan 2010; Ghisellini 2010). The temporal structures of the prompt
emission are very complicated, consisting of many overlapping
pulses. Pulses are the basic, central building blocks of the prompt
emission, and their correlative properties imply that the pulses are
responsible for many luminosity-related characteristics. Recent
studies showed that the lag vs. luminosity relation (Norris et al.
2000), the variability vs. luminosity relation (Reichart et al.
2001), the $E_{\rm peak}$ vs. $E_{\rm iso}$ relation (Amati et al. 2002) and
the $E_{\rm peak}$ vs. $L_{\rm iso}$ relation (Wei \& Gao 2003; Yonetoku et al. 2004) all
seem to be better explained by pulse rather than bulk emission properties
(see, Hakkila et al. 2008; Hakkila \& Cumbee 2009; Krimm et al.
2009; Firmani et al. 2009; Ohno et al. 2009; Ghirlanda, Nava \&
Ghisellini 2010; Arimoto et al. 2010). In principle, the bulk characteristics of the
prompt emission can be derived from our knowledge of the decomposition of the burst
in pulses and their individual properties. Therefore, it is essential to our
understanding of the physics of the bulk prompt emission of GRBs, that we properly
measure and understand the properties of the individual pulses.

Hakkila et al. (2008) isolated and delineated pulse spectral
properties of GRBs detected by BATSE with known redshifts, and found that
pulse lag, pulse luminosity, and pulse duration are strongly
correlated. They also found that pulse peak lag, pulse asymmetry, and
pulse hardness are correlated for a large number of pulses of long GRBs
(Hakkila \& Cumbee 2009). These results indicate that most
pulses of long GRBs within a given burst as well as when comparing different bursts might
have similar physical origins.

However, in some cases, which show two or more separated distinct
emission episodes, and each emission episode consists of one or more
pulses, their pulse properties and origins are likely complicated.
For example, Hakkila \& Giblin (2004) identified two cases
(GRBs 960530 and 980125), consisting of two separated emission episodes,
and found that the pulses in the second emission episodes of these two GRBs have
longer lags, smoother morphologies, and softer spectral evolution
than those in the first episodes. It has been suggested that
internal- and external-shock emission might overlap in these two
cases (Hakkila \& Giblin 2004).

Recently, the Swift Burst Alert Telescope (BAT) detected a burst,
GRB 090618, which shows two emission episodes with four prominent
pulses (Baumgartner et al. 2009). It is obvious that the pulse in
the first episode has a smoother morphology than the three pulses in
the second episode. We wonder whether the pulses in the two emission
episodes within this burst have different properties and/or origins.
To this end, we have performed a detailed analysis of the pulse
temporal and spectral characteristics of GRB 090618 (preliminary results are
reported in Zhang 2011).

\section{Observations}

GRB 090618 was detected by Swift-BAT at 08:28:29 UT on 2009 June 18
(this time is used as $T_{0}$ throughout the paper, Schady et al.
2009). The burst was also observed by Fermi-GBM (McBreen et al.
2009), AGILE (Longo et al. 2009), Suzaku WAM (Kono et al. 2009),
KONUS-WIND and KONUS-RF (Golenetskii et al. 2009). The Swift X-ray
telescope (XRT) began follow up observations of its X-ray light
curve 124 s after the BAT trigger and its UVO telescope detected its
optical afterglow 129 s after the trigger (Schady et al. 2009).
Absorption features which were detected in its bright optical
afterglow with the 3m Shane telescope at Lick observatory yielded a
redshift of $z=0.54$ (Cenko et al. 2009).

The BAT burst light curve shows a smooth multipeak structure with 4
prominent pulses. Significant spectral evolution was observed during the burst. The spectrum at
the maximum count rate, measured from $T_{0}$+62.720 to $T_{0}+64.0$
s, was well fitted (Golenetskii et al. 2009) in the 20 keV$-$2 MeV
range by the Band function (Band et al. 1993) with a low-energy
photon index $-0.99 (-0.06, +0.07)$, a high energy photon index
$-2.29 (-0.5, +0.23)$, and peak energy $E_{p}=440\pm70$ keV, while
the time integrated spectrum had a low-energy photon index
$-1.28\pm0.02$, a high energy photon index $-2.66(-0.2, +0.14)$, and
a peak energy $E_{p}=186\pm8$ keV (Golenetskii et al. 2009). The
isotropic equivalent energy in the 8$-$1000 keV band was
$E_{iso}=2.0\times10^{53}$ erg (standard cosmology, McBreen et al.
2009).

\section{Pulse Temporal Properties}

Figure 1 shows the BAT and GBM light curves over the standard energy
bins (BAT: 15$-$25, 25$-$50, 50$-$100 and 100$-$350 keV; GBM:
8$-$1000 keV (NaI) and 0.2$-$30 MeV (BGO)). The first episode
(episode A) is a smooth 50 s pulse starting at $T_{0} - 5$ s, and
ends at $T_{0} + 45$ s (pulse A). The second episode (episode B)
starts at $\sim T_{0} + 45$ s and is about 275 s long, consisting of
three overlapping pulses. The first pulse peak at $\sim T_{0} + 62$
s (pulse B1), the second peak is at $\sim T_{0} + 80$ s (pulse B2),
and the third peak is at $T_{0} + 112$ s, finally ending at $T_{0} +
320$ s (pulse B3). $T_{90}$ (15$-$350 keV) is $113.2\pm0.6$ s
(estimated error including systematics, Baumgartner et al. 2009). We
focus attention on how the pulse width and pulse width ratio
depend on energy in the two emission episodes, while checking if
that dependence is maintained during this burst.

\begin{figure}
\centering
\includegraphics{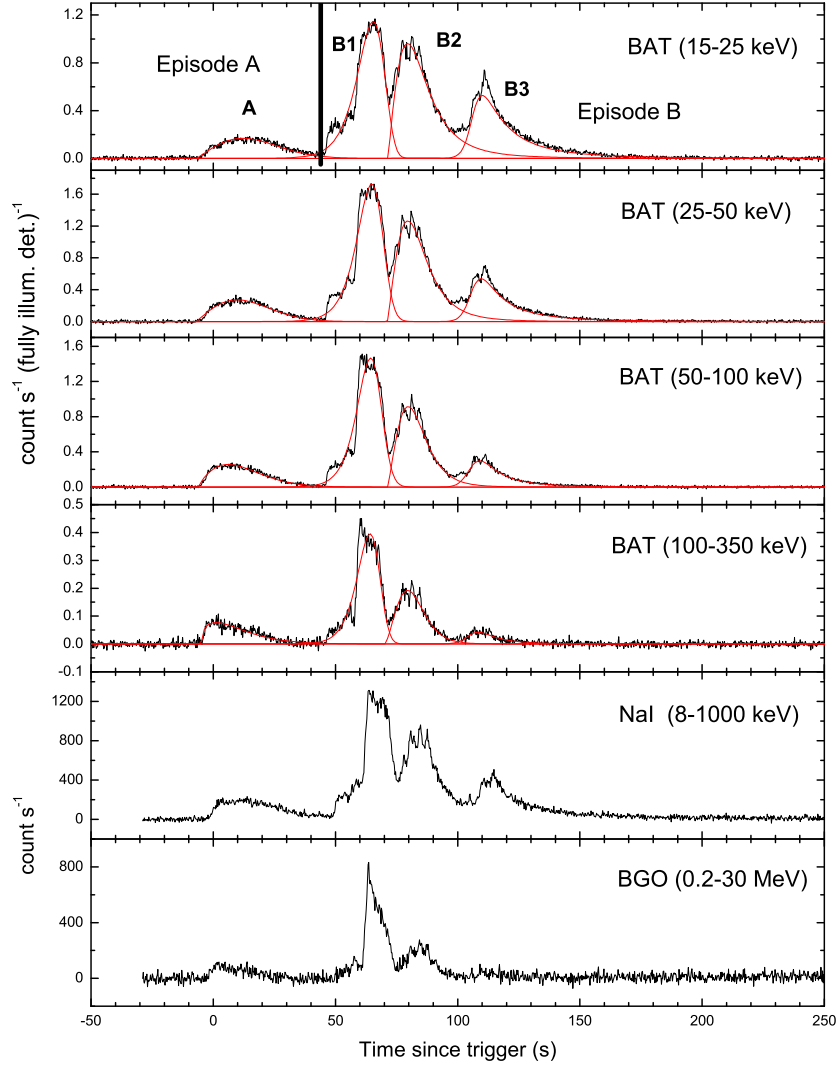}
\caption{Broadband light curves of GRB 090618 observed by Swift and
Fermi. The fitting curves with eq. (1) are plotted. \label{fig1}}
\end{figure}

Kocevski et al. (2003) developed an empirical expression, which can
be used to fit the pulses of GRBs. This function can be written
as,
\begin{equation}
F(t)=F_{m}(\frac{t+t_{0}}{t_{m}+t_{0}})^{r}[\frac{d}{d+r}+\frac{r}{d+r}
(\frac{t+t_{0}}{t_{m}+t_{0}})^{(r+1)}]^{-\frac{r+d}{r+1}},
\end{equation}
where $t_{m}$ is the time of the maximum flux ($F_{m}$) of the
pulse, $t_{0}$ is the offset time, $r$ and $d$ are the rising and
decaying power-law indices, respectively. Because the prompt emission
of GRB 090618 is concentrated mainly in the Swift-BAT energy range, only the BAT light
curves are considered. We fit all the light curves (see Figure
1) of the burst in the different BAT energy bands with equation (1)
and then measure the values of pulse-width ($w$) and the
rise-to-decay ratio of pulse ($r/d$, pulse asymmetry). The errors of $w$ and
$r/d$ are derived from simulations by assuming a normal
distribution of the errors of the fitting parameters. The reported
errors are at $1\sigma$ confidence level. The results are listed in
Table 1.

\begin{table*}
 \begin{minipage}{100mm}
  \caption{Pulse temporal characteristics of GRB 090618.}
 \centering
  \begin{tabular}{lcccccccc}
  \hline
       &\multicolumn{2}{c}{Pulse A} &\multicolumn{2}{c}{Pulse B1} &\multicolumn{2}{c}{Pulse B2} &\multicolumn{2}{c}{Pulse B3%
       }\\
 Band & $w$     & $r/d$   & $w$       & $r/d$ & $w$        & $r/d$&  $w$        & $r/d$ \\
 (keV) & (s) & & (s) & & (s) & & (s) & \\

\hline
 (1) 15-25    &32.3$\pm$4.6 &0.68$\pm0.11$ &13.2$\pm1.4$ &1.44$\pm0.29$ &16.1$\pm1.4$ &0.55$\pm0.11$ &15.5$\pm3.5$ &0.48$\pm0.13$ \\
 (2) 25-50    &28.4$\pm$3.2 &0.63$\pm0.09$ &12.6$\pm1.1$ &1.36$\pm0.29$ &15.1$\pm1.5$ &0.59$\pm0.13$ &15.0$\pm4.2$ &0.52$\pm0.16$ \\
 (3) 50-100   &24.9$\pm$2.4 &0.57$\pm0.10$ &11.9$\pm1.1$ &1.29$\pm0.27$ &14.1$\pm2.1$ &0.65$\pm0.15$ &14.3$\pm7.2$ &0.59$\pm0.24$ \\
 (4) 100-350  &20.3$\pm$5.1 &0.37$\pm0.10$ &11.1$\pm1.8$ &1.31$\pm0.35$ &13.3$\pm3.1$ &0.71$\pm0.19$ &13.4$\pm6.9$ &0.62$\pm0.26$ \\
\hline
\end{tabular}
\end{minipage}
\end{table*}

From Table 1, we find a significant trend: all the pulses
tend to be narrower at higher energies. However, the pulse
asymmetry dependence on the energy are different for the two
emission episodes. The pulses B2 and B3 tend to be more symmetric at
higher energies while the pulse A follows a reverse trend. To
further study how the pulse width depends on energy in detail, we
show $w$ and $r/d$ as functions of energy ($E$) in Figure 2, where
$E$ is the geometric mean of the lower and upper boundaries of the
corresponding energy band (this is adopted throughout this paper
unless otherwise noted). Apparently both $w$ and $r/d$ are correlated
with $E$. The correlation analysis yields $w\propto
E^{-0.20\pm0.01}$ and $r/d\propto E^{-0.24\pm0.06}$ for the pulse A,
$w\propto E^{-0.07\pm0.01}$ and $r/d\propto E^{-0.05\pm0.04}$ for
the pulse B1, $w\propto E^{-0.09\pm0.01}$ and $r/d\propto
E^{0.12\pm0.01}$ for the pulse B2, and $w\propto E^{-0.06\pm0.01}$
and $r/d\propto E^{0.12\pm0.02}$ for the pulse B3. It is found that
the $w - E$ relations of GRB 090618 are well consistent with those observed in the
majority of long GRBs (e.g., Norris et al. 1996; 2005; Peng et al. 2006),
but the power-law indices of the $w - E$ relations within this event
are larger than those previously observed in typical
GRBs (e.g., Fenimore et al. 1995; Norris et al. 1996; 2005), and the
indices in the episode B are larger than that in the episode A.
The large power-law indices of the $w - E$ relations in GRB 090618 can be
explained from the fact that the distribution of power-law
index of the $w - E$ relation has a large dispersion (see, Jia \&
Qin 2005; Peng et al. 2006; Zhang et al. 2007, Zhang \& Qin 2008;
Zhang 2008). In addition, we also find that the energy dependence
of $r/d$ is different for the 4 pulses in the burst.
The power-law indices of $r/d - E$ relation for the pulses A and B1
are negative\footnote{For the pulse B1, the power-law
anti-correlation between $r/d$ and $E$ is not very robust, this is
so because the pulse rising phase is likely affected by overlapping
mini-pulses (see Figure 1).}, while the power-law indices of the relation for the pulses
B2 and B3 are positive. The two different energy dependence correlations of $r/d$
were observed previously within different bursts for a large set of GRBs in the BATSE
database (see, Peng et al. 2006). The power-law correlation between
$r/d$ and $E$ has been predicted theoretically by Qin et al (2004; 2005), who
suggested that the emission associated with the shocks
occurs on a relativistically expanding fireball surface, where
the curvature effect must be important. However, it is unclear which
mechanism is responsible of the power-law anti-correlation between $r/d$ and $E$.
As proposed by Peng et al. (2006), a varying synchrotron or
comptonized radiation or a different pattern of the spectral
evolution should be considered. Furthermore, the different dependence on energy of
pulse asymmetries in one single GRB is reported firstly, this indicates
that the evolution and/or nature of pulses might different in some GRBs and
the different emission episodes are likely to originate from different physical
mechanisms (e.g., Hakkila \& Giblin 2004).

\begin{figure}
\centering
\includegraphics[]{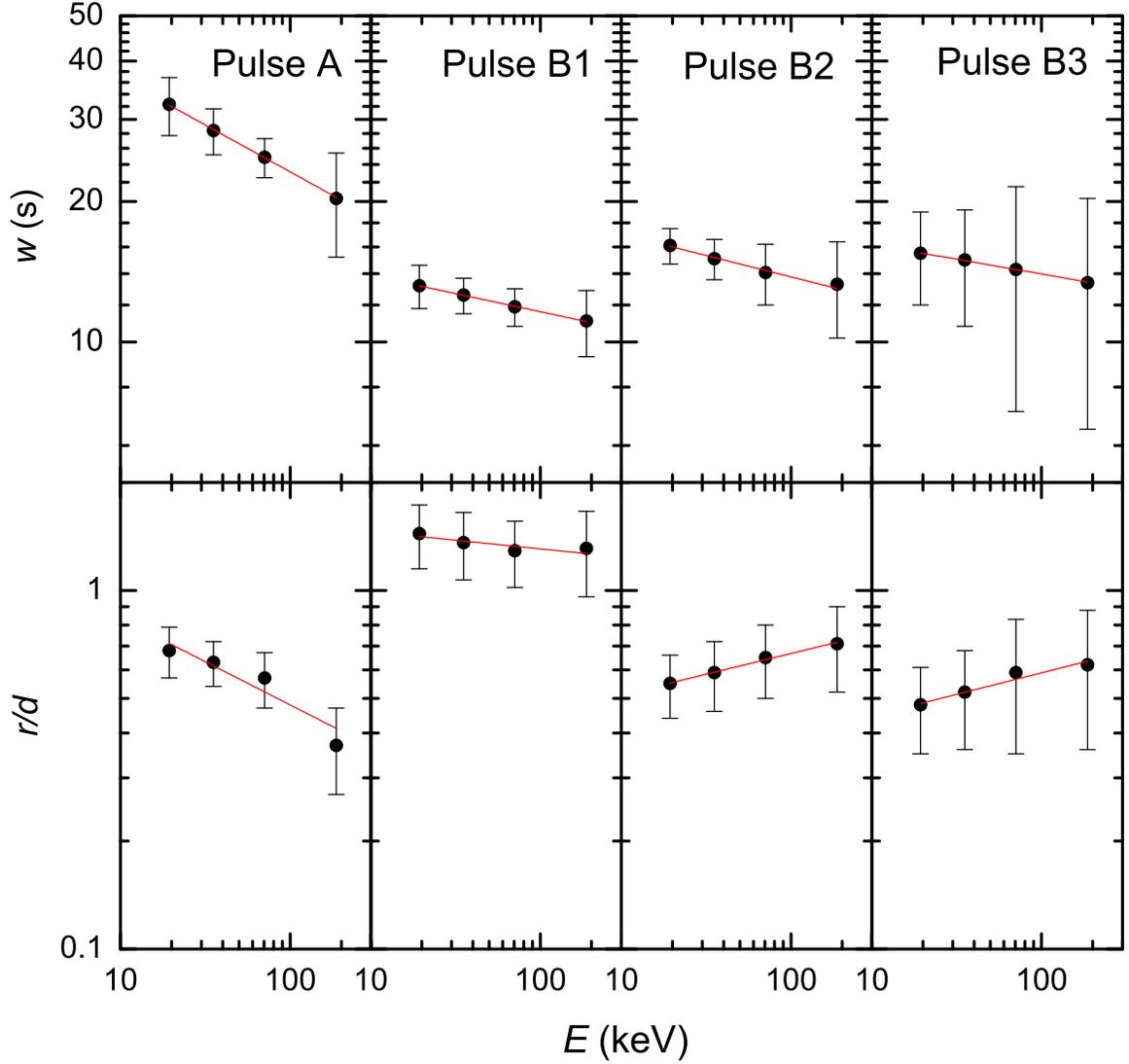}
 \caption{Dependence of the pulse width
($w$, $top$ $panels$) and pulse rise-to-decay ratio ($r/d$, $bottom$
$pannels$) on energy in GRB 090618. The solid lines in the plots represent
the best fits. \label{fig2}}
\end{figure}

\section{Pulse Spectral characteristics}

To further check if the pulses in the two emission episodes of GRB
090618 have different properties and/or different physical origins,
we have performed a detailed pulse spectral analysis.

\subsection{Pulse spectral evolution}

Pulse spectral evolution is very important to understand the
physics of GRB pulses (and thus of GRB prompt emission).
Page et al. (2011) performed 14 time-slices spectra for GRB 090618,
and found that the peak energy initially decreases with time, then moves
to higher energies during flaring activity. In general,
there is a positive trend between peak energy and flux.
In order to perform a more detailed study of the individual pulse spectral evolution in GRB 090618, 23
time-sliced spectra from both the Swift-BAT and Fermi-GBM (NaI and
BGO) detectors, covering $-5-$150 s after the trigger, were extracted
with single power-law (BAT) and cutoff power-law (CPL, joint
BAT-GBM) models\footnote{The BAT and GBM data are publicly
available at http://swift.gsfc.nasa.gov/ and
http://fermi.gsfc.nasa.gov/.}. In general, 5 s time interval is selected to perform
time-resolved spectral analysis. For the weak emission in the begin and end stage of pulses,
10 s or 30 s time interval is adopted (see Table 2). The standard data
analysis methods according to the BAT Analysis
Threads\footnote{http://heasarc.gsfc.nasa.gov/docs/swift/analysis/threads/
bat\_threads.html} and the GBM Analysis
Threads\footnote{http://fermi.gsfc.nasa.gov/ssc/data/analysis/scitools/
gbm\_grb\_analysis.html} are used.
The useful energy ranges for the BAT, NaI
and BGO spectral fitting are 15$-$150, 8$-$1000 and 200$-$30 000
keV, respectively. Spectra were analyzed with Xspec(v12) software.
Note that the Band model (Band et al. 1993) is extensively used to
fit the GRB spectra. For GRB 090618, the high energy index in the Band model
cannot be well constrained in most of the time
slices (also see Page et al. 2011). For the purpose of comparing the spectral evolution of
the different pulses of GRB 090618 under one spectral model, we choose the minimal simplest model, i.e. the
CPL.  The power-law index ($\Gamma_{\rm PL}$) from the BAT fit with the single PL model, and the
peak energy $E_{\rm peak}$ and low-energy index ($\Gamma_{\rm CPL}$) from the
joint BAT-GBM fits with the CPL model are shown in Figure 3 and Table 2.

\begin{table*}
 \begin{minipage}{140mm}
  \caption{Spectral results of the time resolved
analysis in GRB 090618.} \centering
 \begin{tabular}{llccccc}

  \hline
      & & \multicolumn{2}{c}{PL (BAT)} &\multicolumn{3}{c}{CPL (BAT+GBM)}%
       \\
t$_{1}$ & t$_{2}$ &$\Gamma_{\rm PL}$ & $\chi^{2}$/dof & $\Gamma_{\rm CPL}$ & E$_{\rm peak}$ & $\chi^{2}$/dof \\
s & s & &  & & keV & \\
\hline

-5  &   5   &   1.01    $\pm$   0.02    &   38/56   &   0.71    $\pm$   0.03    &   235 $\pm$   15  &   432/410 \\
5   &   10  &   1.26    $\pm$   0.03    &   62/56   &   0.89    $\pm$   0.04    &   193 $\pm$   18  &   381/410 \\
10  &   15  &   1.38    $\pm$   0.03    &   63/56   &   0.97    $\pm$   0.06    &   156 $\pm$   19  &   418/410 \\
15  &   20  &   1.51    $\pm$   0.03    &   64/56   &   1.11    $\pm$   0.07    &   155 $\pm$   24  &   467/410 \\
20  &   25  &   1.66    $\pm$   0.03    &   77/56   &   1.3 $\pm$   0.09    &   162 $\pm$   39  &   472/410 \\
25  &   35  &   1.87    $\pm$   0.04    &   77/56   &   1.05    $\pm$   0.16    &   64  $\pm$   13  &   454/410 \\
35  &   45  &   2.16    $\pm$   0.08    &   61/56   &   1.53    $\pm$   0.22    &   98  $\pm$   55  &   434/410 \\
45  &   50  &   1.69    $\pm$   0.03    &   43/56   &   1.37    $\pm$   0.04    &   317 $\pm$   50  &   463/410 \\
50  &   55  &   1.52    $\pm$   0.02    &   38/56   &   1.2 $\pm$   0.03    &   313 $\pm$   28  &   524/410 \\
55  &   60  &   1.41    $\pm$   0.02    &   39/56   &   1.06    $\pm$   0.1 &   500 $\pm$   26  &   922/410 \\
60  &   65  &   1.38    $\pm$   0.01    &   44/56   &   1.14    $\pm$   0.1 &   389 $\pm$   13  &   721/410 \\
65  &   70  &   1.51    $\pm$   0.01    &   53/56   &   1.23    $\pm$   0.02    &   234 $\pm$   12  &   508/410 \\
70  &   75  &   1.74    $\pm$   0.02    &   56/56   &   1.37    $\pm$   0.02    &   245 $\pm$   17  &   604/410 \\
75  &   80  &   1.66    $\pm$   0.02    &   50/56   &   1.31    $\pm$   0.02    &   278 $\pm$   14  &   642/410 \\
80  &   85  &   1.6 $\pm$   0.02    &   57/56   &   1.3 $\pm$   0.02    &   250 $\pm$   14  &   577/410 \\
85  &   90  &   1.71    $\pm$   0.02    &   60/56   &   1.33    $\pm$   0.04    &   153 $\pm$   14  &   455/410 \\
90  &   95  &   1.83    $\pm$   0.02    &   63/56   &   1.39    $\pm$   0.07    &   116 $\pm$   15  &   421/410 \\
95  &   100 &   1.99    $\pm$   0.03    &   60/56   &   1.39    $\pm$   0.09    &   81  $\pm$   12  &   415/410 \\
100 &   105 &   1.97    $\pm$   0.03    &   78/56   &   1.4 $\pm$   0.07    &   103 $\pm$   13  &   449/410 \\
105 &   110 &   1.97    $\pm$   0.02    &   73/56   &   1.48    $\pm$   0.05    &   111 $\pm$   10  &   409/410 \\
110 &   115 &   2.13    $\pm$   0.02    &   63/56   &   1.6 $\pm$   0.06    &   97  $\pm$   11  &   437/410 \\
115 &   120 &   2.22    $\pm$   0.03    &   70/56   &   1.56    $\pm$   0.07    &   74  $\pm$   10  &   395/410 \\
120 &   150 &   2.39    $\pm$   0.03    &   61/56   &   1.81    $\pm$   0.1 &   80  $\pm$   14  &   447/410 \\

  \hline
\end{tabular}
\end{minipage}
\end{table*}

\begin{figure}
\centering
\includegraphics[]{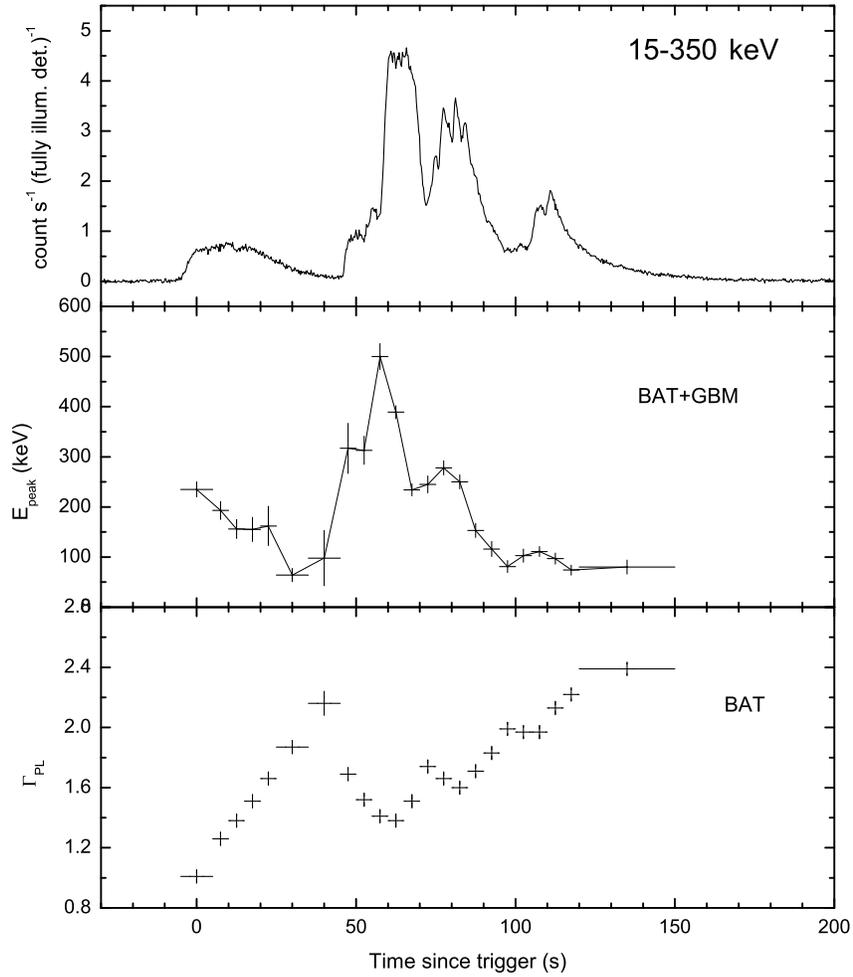}
\caption{Spectral evolution of GRB 090618, where the values of
$E_{\rm peak}$ are obtained from the joint GBM-BAT fits with the cutoff
power-law model, and the values of $\Gamma_{\rm PL}$ are measured from
the Swift-BAT fit with the single power-law model. The light curve
in the BAT band ($15-350$ keV) is also displayed.\label{fig3}}
\end{figure}

From Figure 3, we conclude that GRB 090618 exhibits significant
spectral evolution and the pulses in the different episodes have
different spectral evolution trends\footnote{Here we only show the
whole spectral evolution of GRB 090618 to depict individual pulse spectral evolution.
It is known that the individual pulse spectrum cannot be divided from a GRB which
have several overlapping pulses. The three pulses in the episode B of GRB 090618 are overlapping,
but they can be identified well (see Figure 1).
Therefore, the individual pulse spectral evolution trend in the episode B cannot be
significantly affected by the overlapping effect.}. $E_{\rm peak}$ of the pulse A shows a
hard-to-soft evolutionary pattern, decreasing monotonically while
the flux rises and falls, $\Gamma$ shows an opposite trend. In the
three pulses of the episode B, there is a positive trend between $E_{\rm peak}$
and flux, while $\Gamma$ follows an opposite trend. The two
types of spectral evolution patterns have been previously observed
in pulses from different GRBs (e.g., Golenetskii et
al.1983; Norris et al. 1986; Preece et al. 1998; Kaneko et al.
2006), but the phenomenon that the two types of spectral evolution
patterns exist simultaneously in one single GRB
is very infrequent. GRB 921207 is another case following such
spectral evolution trend (see, Figure 4 of Ford et al. 1995 and
Figure 2 of Lu et al. 2010). It is difficult to accommodate the two
different spectral evolution trends under one mechanism. Lu et al.
(2010) argued that it could be explained in terms of the viewing angle and
jet structure effects.

\subsection{Pulse spectral lag}

\begin{table*}
 \begin{minipage}{140mm}
 \centering
  \caption{Pulse peak lags of GRB 090618. The numbering represent the energy bands
  used to calculated the pulse peak lags listed in the column 1 of Table 1 (e.g., Lag 21
  represent the lag is measured between (2)25-50 keV and (1)15-25 keV energy bands.).}

  \begin{tabular}{lllllll}
  \hline

Pulse & Lag 21& Lag 31& Lag 41& Lag 32 & Lag 42&Lag 43\\
  & (s)&  (s) &  (s)& (s)& (s)& (s) \\
\hline
 Pulse A    &2.19$\pm0.36$ &6.01$\pm0.37$ &10.88$\pm0.36$ &3.82$\pm0.12$ &8.69$\pm0.15$ &4.87$\pm0.20$ \\
 Pulse B1   &0.53$\pm0.20$ &0.98$\pm0.21$ &1.34$\pm0.08$  &0.45$\pm0.09$ &0.81$\pm0.10$ &0.36$\pm0.20$ \\
 Pulse B2   &0.32$\pm0.20$ &-0.20$\pm0.20$ &0.38$\pm0.22$  &-0.52$\pm0.19$ &0.06$\pm0.13$ &0.58$\pm0.14$  \\
 Pulse B3   &0.49$\pm0.56$ &1.06$\pm0.56$ &1.70$\pm0.75$  &0.57$\pm0.21$ &1.21$\pm0.54$ &0.64$\pm0.54$  \\
\hline

\end{tabular}
\end{minipage}
\end{table*}

Another observed effect of the spectral evolution in GRB data is spectral lag.
Spectral lags are energy-dependent delays in the GRB temporal
structure. Pulse peak lags are defined as the differences
between the pulse peak times in different energy channels, which can
be obtained for any pulse between two energy channels (e.g., Norris
et al. 2005; Liang et al. 2006; Hakkila et al. 2008; Zhang et al.
2007; Zhang 2008). In general, soft pulses lag behind hard pulses. The pulse
peak-fit method gives a simple straightforward way for extracting
lags ( Norris et al. 2005; Hakkila et al. 2008). The pulse spectral lags
between the four standard BAT energy bands (see Table 1) are displayed in Table 3.
We find that the pulse A of GRB 090618 has a
very longer lag (in all energy channel combinations) than
all three pulses in the episode B. Using the cross-correlation function (CCF)
analysis method, Page et al. (2011) analyzed the whole spectral
lags in the two episodes of GRB 090618 and found that the episode A have a
lag about a factor of 6 longer than for the episode B. Their result is consistent with
our finding, although the episode B of GRB 090618 comprises three pulses.
A similar phenomenon was also obtained by Hakkila \& Giblin (2004). The
early studies of burst spectral lags show that they vary within a
given burst as well as from burst to burst (e.g., Norris 2002; Ryde
et al. 2005; Chen et al. 2005). Multi-lag GRBs are ubiquitous.
Therefore, we can not differentiate between their physical origins by only
taking into account the spectral lags.

\section{Conclusions and Discussion}

\begin{figure}
\centering
\includegraphics{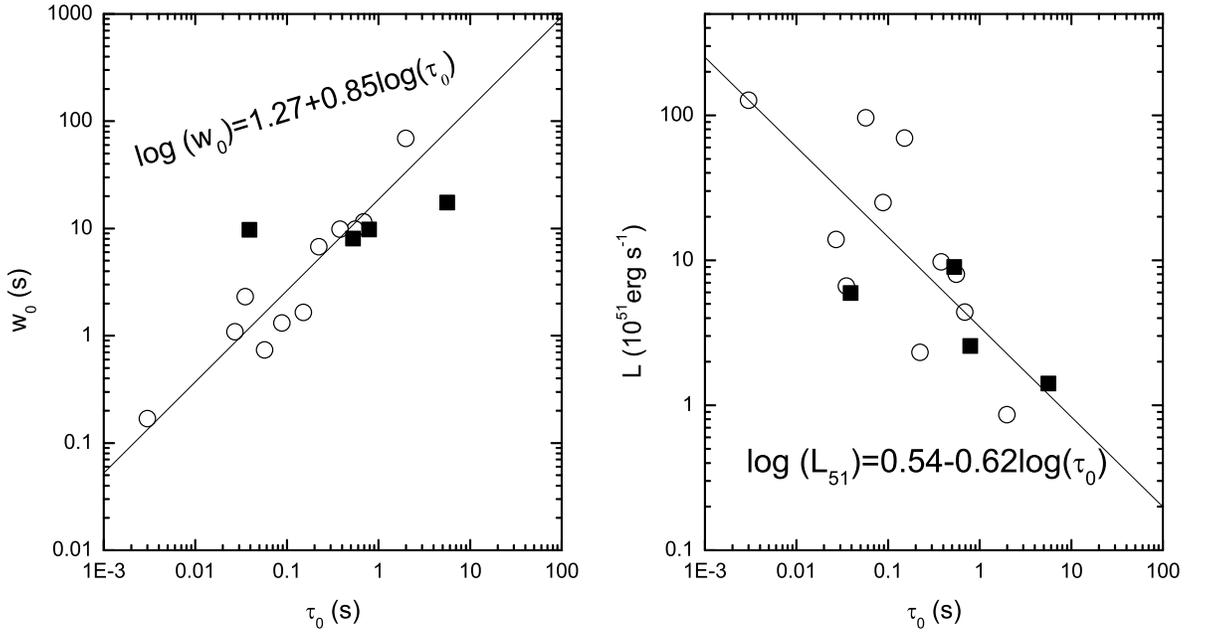}
 \caption{\emph{Left}: Rest frame pulse
duration $w_{0}$ vs. pulse peak lag $\tau_{0}$ for fit pulses of
BATSE GRBs having known redshifts (the data are taken from Hakkila
et al. 2008, GRB 980425 is excluded) as well as GRB 090618.
\emph{Right}: Isotropic pulse peak luminosity $L$ vs. pulse peak lag
$\tau_{0}$ for the pulses shown in the left panel. The open circles
represent the pulses from GRB 971214, GRB 980703, GRB 970508, GRB
990510, GRB 991216 and GRB 990123, and the filled square represent
GRB 090618. The solid lines are the best fits obtained by Hakkila et
al. (2008). \label{fig4}}
\end{figure}

In this work we have used the pulse peak-fit method to analyze the
pulse temporal and spectral characteristics of GRB 090618. We
find that the pulses in the two emission episodes have different
properties, including the energy dependence of pulse widths and the
pulse asymmetries, the pulse spectral evolution patterns as well as
the pulse lags. The different pulse temporal and spectral
characteristics exhibit simultaneously in one single GRB, indicating
there might be different origins in the different emission episodes of some GRBs. None of the
mechanisms proposed so far can be used to account for this fact. Recently, Hakkila et al.
(2008) and Hakkila and Cumbee (2009) found that isotropic pulse peak
luminosity ($L$), rest frame pulse peak lag ($\tau_{0}$), and pulse
duration ($w_{0}$) are correlated intrinsic properties of most GRB
pulses, and argued that most pulses might result from variations
on a single pulse type. To further understand the different pulse
properties, we also calculated the values of $L$, $\tau_{0}$ and
$w_{0}$ for all pulses in GRB 090618 and compare their relations
with the Hakkila et al. (2008) result (Figure 4). We find that the distributions of $L$, $\tau_{0}$ and $w_{0}$ for
the four pulses basically comply with the relations found by
Hakkila et al. (2008). Such a result renders the interpretation of the different pulse temporal and spectral properties found in our earlier analysis much more challenging.
The first episode is dimmer than the second episode and may be identified as a precursor of the burst.
A precursor could either have the
same origin as the main emission episode or it could be due to
a different mechanism (see, Koshut et al. 1995; Lazzati 2005; Burlon et
al. 2008, 2009). Recently, Penacchioni et al. (2011) proposed that GRB 090618 might
be a members of a specific new family of GRBs presenting a double astrophysical component.
A first one, related to the proto-black hole, prior to the process of gravitational collapse (episode A)
and a second one which is the canonical GRB (episode B) emitted during the formation of the black-hole.

Better measurements are needed in order to improve our understanding of GRB pulse
properties. Description and analysis of pulse properties can help
to constrain physical models. The similar time evolution of
pulse structures, combined with the fact that their measurable
properties correlate strongly, suggests that one physical mechanism
produces the observed array of pulse characteristics (see Hakkila et al.
2008; Hakkila \& Cumbee 2009). There is strong evidence that the
majority of GRB pulses results from internal shocks in relativistic
winds (e.g. Sari \& Piran 1997; Kobayashi et al. 1997; Daigne \&
Mochkovitch 1998; Ramirez-Ruiz \& Fenimore 2000; Nakar \& Piran
2002). Katz (1994) suggested that GRB pulse shapes originate from
time delays inherent in the geometry of spherically expanding
emission fronts. Liang et al. (1997) argued that saturated Compton
up-scattering of softer photons may be the dominant physical
mechanism that creates the shape of GRB pulses. According to Ryde
and Petrosian (2002), the simplest scenario accounting for the
observed GRB pulses is to assume an impulsive heating of the leptons
and their subsequent cooling and emission. In addition, in the impulsive
external shock model, a single relativistic wave of plasma interacts
with inhomogeneities in the surrounding medium and form external
shocks that accelerate particles which can also contribute to the
formation of GRB pulses (Dermer et al. 1999). Although many studies have been performed to
interpret the pulse characteristics, their nature is still unclear.

\textbf{Acknowledgments}
We acknowledge the use of the public data from the Swift and Fermi
data archive. We thank the anonymous referee for very detailed comments
which helps in improving the quality of the paper. We also thank Yi-Zhong Fan for insightful comments.
This work was supported by the National Natural Science Foundation of China (grants 11163003)
and Guangxi Natural Science Foundation (2010GXNSFB013050).

\end{document}